\documentclass[runningheads]{llncs}
\usepackage{graphicx}

\usepackage{times}
\usepackage{latexsym}

\usepackage[T1]{fontenc}

\usepackage[utf8]{inputenc}
\usepackage{csquotes}

\usepackage{amsfonts}
\usepackage{microtype}
\usepackage[column=O]{cellspace}
\usepackage{booktabs}
\usepackage{adjustbox}
\usepackage{tabularx}
\usepackage[none]{hyphenat}
\usepackage{lipsum}
\usepackage{amsmath,amssymb}
\usepackage{multirow}
\usepackage{capt-of}
\usepackage[colorinlistoftodos]{todonotes}
\usepackage{caption} 
\usepackage{float}
\usepackage{longtable}
\usepackage[bb=boondox]{mathalfa}
\usepackage{authblk}
\usepackage[misc]{ifsym}
\usepackage[
    backend=biber,
    style=nature,
  ]{biblatex}
\usepackage{algorithm}
\usepackage{algpseudocode}
\usepackage{verbatim}

\usepackage{subfig}
\let\llncssubparagraph\subparagraph
\let\subparagraph\paragraph
\usepackage[compact]{titlesec}
\let\subparagraph\llncssubparagraph
\titlespacing{\section}{0pt}{2ex}{1ex}
\titlespacing{\subsection}{0pt}{1ex}{0ex}

\begin{document}
\title{Improving Document Image Understanding with Reinforcement Finetuning}
\author{Bao-Sinh Nguyen\inst{1}\Letter \and Dung Tien Le\inst{1} \and Hieu M. Vu\inst{1} \and Tuan-Anh D. Nguyen\inst{1} \and \quad \quad \quad   Minh-Tien Nguyen\inst{1,3} \and Hung Le\inst{2}}

\authorrunning{Nguyen et al.}


\institute{Cinnamon AI \\10th floor, Geleximco building, 36 Hoang Cau, Dong Da, Hanoi, Vietnam. \\ \email{\{simon, nathan, ian, tadashi, ryan.nguyen\}@cinnamon.is} \and
    Deakin University, Australia.\\
    \email{thai.le@deakin.edu.au}\\
    \and
    Hung Yen University of Technology and Education, Vietnam.\\
    \email{tiennm@utehy.edu.vn}}

\toctitle{Improving Document Image Understanding with Reinforcement Finetuning}
\tocauthor{Bao-Sinh Nguyen, Dung Tien Le, Hieu M. Vu, Tuan-Anh D. Nguyen, Minh-Tien Nguyen, Hung Le}

\maketitle

\begin{abstract}
    Successful Artificial Intelligence systems often require numerous labeled data to extract information from document images. In this paper, we investigate the problem of improving the performance of Artificial Intelligence systems in understanding document images, especially in cases where training data is limited. We address the problem by proposing a novel finetuning method using reinforcement learning. Our approach treats the Information Extraction model as a policy network and uses policy gradient training to update the model to maximize combined reward functions that complement the traditional cross-entropy losses. Our experiments on four datasets using labels and expert feedback demonstrate that our finetuning mechanism consistently improves the performance of a state-of-the-art information extractor, especially in the small training data regime. 
    \keywords{Information Extraction \and Reinforcement Learning \and Human-In-The-Loop}
\end{abstract}

\section{Introduction}
Digitizing business documents is crucial for companies and corporations to improve their productivity and efficiency. Although the advent of Document Intelligence brings forth many opportunities to capture the key information of document images, extraction for visually-rich documents such as receipts, invoices, and leaflets remains notoriously challenging due to the spareness of textual information and the variety in layouts and formats. Thus, for an AI system to fully understand and extract desired information, it is essential to incorporate the textual, visual, and layout aspects into the model and have it trained in an end-to-end manner.

To this end, several approaches have aimed to encode visual and layout information in addition to contextual features to enhance document representations. Earlier approaches focus on improving the performance of document layout and information extraction separately, and combine them in a multi-stage pipeline. However, these methods often lead to cascading error due to the nature of pipeline processing. Recent attempts introduce end-to-end training via Graph Neural Networks \cite{liu2019gcn} and language modeling - LayoutLM \cite{xu2020layoutlm,xu2021layoutlmv2} - for jointly modeling text and layout information.

While the methods above are effective when dealing with benchmark datasets, there exists challenges for practical cases. First, for the model to perform well on a new dataset, a decent amount of labeled training data are required to finetune the model. Benchmark datasets can include up to thousand samples or even more. However, in reality, only a few labeled samples can be provided for training because data annotation is time-consuming and labor-expensive. Therefore, an open research question still remains: \textit{how do we make the most of these valuable data to enable sample-efficient information extraction?} The second issue is the current training process relies solely on predefined datasets for finetuning in a supervised manner. The supervised training often uses the differentiable loss, such as cross-entropy, to predict the locations of the extracted content,  which may cause a mismatch between the training objective and the performance metric. 
More importantly, this opts out the opportunities for domain experts to provide feedback, design the learning criteria, and guide the model to more accurate extraction. 

In this paper, we introduce a new fine-tuning method that complements the traditional cross-entropy training with different learning objectives that match better the evaluation criteria. We formulate the information extraction task as a reinforcement learning (RL) problem wherein the information extractor, such as SpanIE-Recur \cite{nguyen2021span}, is the policy network, and its output corresponds to actions. We design different reward functions to capture the spatial, categorical, lexical and semantic similarity between the extracted and the ground-truth answers. We then use the proximal policy gradient algorithm (PPO) \cite{schulman2017proximal} to learn the optimal policy that maximizes the total rewards. 
Our approach not only exploits all aspects of training signals from the data but also allows the domain expert to design rewards or provide feedback to improve the system. 
We evaluate our method on two public and two private visually-rich document datasets and achieve consistent improvement over only using supervised training in terms of F1 scores, especially when the the training data is limited. 
We also conduct human-in-the-loop experiment where a domain expert provide ranking feedback to finetune the model, showing clear improvement over few number of interactions.   

Our contributions are three-fold: (i) a novel approach to improve information extraction in business documents using reinforcement learning, (ii) a set of complimentary reward functions to aid the traditional supervised learning (SL), and (iii) extensive experiments on both public (English) and private real-world dataset (Japanese), confirming the benefit of our approach with both limited ground-truth labels and expert's feedback.  

\section{Related Work}
\subsubsection{Information extraction (IE)} is an increasingly popular task, where the goal is to automatically extract structured information from a given unstructured document. Unlike plain text, visually-rich documents (VRDs) usually contain sparser texts with well-defined layouts and meaningful visual structures, which can previously be addressed using computer vision and graph-based methods \cite{yu2021pick,davis2021fudge}. With the success of BERT \cite{devlin2019bert} and its derivatives, e.g. LayoutLM \cite{xu2020layoutlm,xu2021layoutlmv2}, IE methods for VRDs are seeing a trend of shifting from traditional to NLP techniques, such as sequence labeling and span extraction.

There have been many studies applying span extraction to IE. \cite{li2021span} presented a span-based model for NER, which enables the model to handle the more common sequence labeling approach such as overlapping entities and discontinued entities. One disadvantage of span-based methods is that they usually only extract one answer for one question at a time. To address this issue, \cite{nguyen2021span} and \cite{son2022jointly} share the same question-context interaction mechanism which enables simultaneous extraction for multiple questions, but the former employs a recursive linking technique while the latter proposes combining span extraction with sequence labeling to facilitate multi-value extraction. In this paper, we adapt SpanIE-Recur to our task. However, instead of directly applying the technique, we empower SpanIE-Recur by putting it into an RL framework, which allows SpanIE-Recur to take into account the advantage of transfer learning and human feedback for training. Based on that, we can improve the quality of our extraction model.

\subsubsection{Reinforcement learning} has been applied to text summarization \cite{celikyilmaz2018deep}, dialogue generation \cite{li2016deeprl} and machine translation \cite{wu2018studyrl}. Most of these methods focus on building an RL agent that generates texts in forms of summary, dialogue, and different language based on the reward signal computed directly from the ground truth text and prediction text. Recent attempts push the idea further by adding preference learning into the reward model \cite{nguyen2021robust,stiennon2020learning,nguyen2022make}. 
We extend RL as the technique to our IE task for VRDs. More importantly, to the best of our knowledge, this paper is the very first attempt to apply reinforcement learning for the visually-rich document information extraction task.

\section{Background}

\subsubsection{Information extraction backbone}
We use SpanIE-Recur \cite{nguyen2021span} as the backbone of our model. SpanIE-Recur addresses the IE problem by the Extractive Question Answering (QA) formulation \cite{devlin2019bert}. Concretely, it replaces the sequence labeling head of the original LayoutLM \cite{xu2020layoutlm} by a span prediction head to predict the starting and the ending positions of the answers given an input field/tag (hereinafter referred as a \textit{question}).


In particular, let $D=\{w_0, w_1, ..., w_n\}$ denote the input document context consisting of $n$ input tokens. The pretrained language model converts them into a set of hidden representations $H=\{h_0, h_1, ..., h_n\}$. The $t$-th question $q_t$ is represented by an embedding vector $e_{q_t}$. The query-context interaction module $g$, which is implemented by attention layers, outputs the starting and the ending position of the corresponding answer span: $start_t, end_t = g(H, e_{q_t})$.



\subsubsection{Proximal Policy Optimization}
Proximal Policy Optimization (PPO) \cite{schulman2017proximal} is an on-policy RL algorithm that utilizes the clipped loss function to avoid big changes in the policy update, still guarantees improvements. Let $s_t$ and $a_t$ denote the state and action at timestep $t$ respectively, $\pi_\theta$ denote the policy network. The PPO training objective is:
\begin{equation}\label{eq_ppo}
    J^{PPO}(\theta) = \mathop{\mathbb{E}}[\min\{ra(\theta)\hat{A}_{\theta_{old}}(s_t,a_t),\\
    \mathrm{clip}(ra(\theta),1-\epsilon,1+\epsilon)\hat{A}_{\theta_{old}}(s_t,a_t)\}]
\end{equation}
where $\theta$ is the current policy's parameters, $\hat{A}_{\theta_{old}}(s_t,a_t)$ is the advantage calculated at the old policy parameters $\theta_{old}$ before each updated policy iteration, by using any advantage estimation algorithm to transform the rewards \cite{schulman2016high}, and $ra(\theta) = \frac{\pi_{\theta}(a_t|s_t)}{\pi_{\theta_{old}}(a_t|s_t)}$ is the ratio between the new policy and the old policy. If the ratio $ra$ falls outside the range $1-\epsilon$ and $1+\epsilon$, the advantage function will be clipped.


\section{Method}
\subsection{Problem formulation for reinforcement finetuning}\label{sec:formulation_rl}
\begin{figure}[!t]
    \centering
    \includegraphics[width=0.9\textwidth]{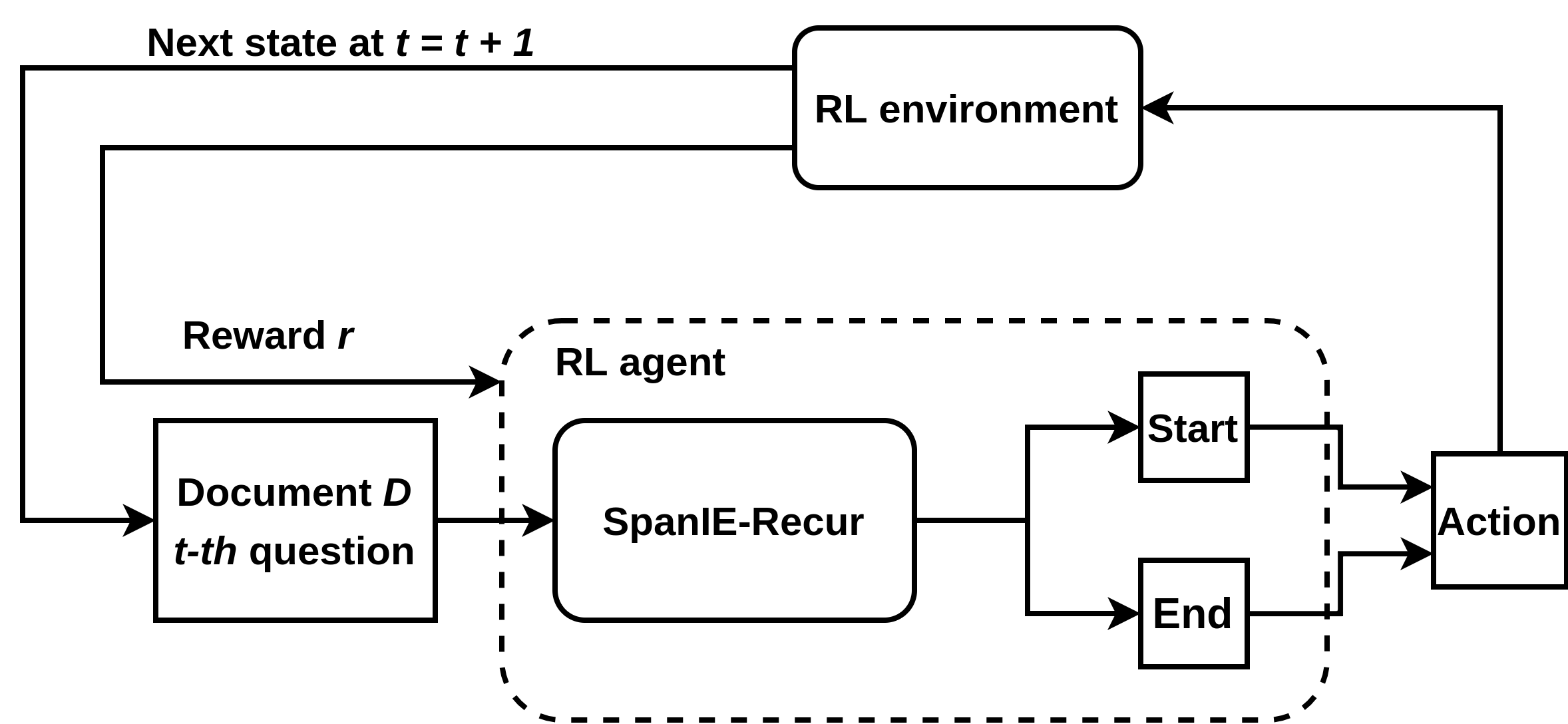}
    \caption{Reinforcement Learning formulation using SpanIE-Recur backbone.}
    \label{fig:spanie-recur}
\end{figure}

We adapt the original SpanIE-Recur \cite{nguyen2021span} as the policy network $\pi_\theta()$ of the IE agent and finetune it using RL. The only difference is that we replace the learnable question embedding $e_{q_t}$ by the embedding produced from a pretrained multilingual text encoder \cite{feng2022language} taking the question text as the input, which can benefit transfer learning settings. We treat the information extraction process as a sequential decision making process so that we can employ RL, where each question to a document corresponds to one timestep in an episode. At each timestep $t$, the document context $D$ and the $t$-th question $q_t$ are fed to the model, that is to say, the agent is in the state $s_t=(D, q_t)$. The model predicts the starting position $start_t$ and the ending position $end_t$ of the answer span, thus the agent takes action $a_t=(start_t, end_t)$, which corresponds to extracting the answer string $\hat{y} = D[start_t:end_t+1]$\footnote{Using Python slicing notation.} from the document context. It then receives a reward $r$ based on a reward function (described in Section \ref{reward}) and goes to the next timestep $t+1$ until all questions are traversed by the agent. The RL formulation is shown in Figure \ref{fig:spanie-recur}.

\subsection{Reward functions}\label{reward}
We improve the quality of SpanIE-Recur by defining new reward functions for RL. We argue that the advantage of RL fine-tuning over SL is that RL fine-tuning provides the flexibility in designing the reward that reflects human’s preferences, which is not available in traditional SL. Moreover, this reward can fully utilize different training signals from the data. Therefore, we introduce an unified reward function, which considers the following criteria of a good answer given an input question.
\subsubsection{String matching reward}
Our string matching reward bases on the Levenshtein distance between the output answer string $\hat{y}$ and the ground truth answer string $y$. Intuitively, we expect that a higher reward should be given to the output answer that is more similar to the ground truth one and vice versa.
    \begin{equation}
    \label{eq_string_rew}
    r_{string} = 1 - \mathrm{Levenshtein\_distance}(y,\hat{y})
    \end{equation}

\subsubsection{Location reward}
The location reward encourages the location matching between the output answer span and the corresponding ground truth span in the document context. Let $a = (start$, $end)$ be the predicted starting and ending positions of the span and $a\_gt = (start\_gt$, $end\_gt)$ be the corresponding ground truth ones. The exact formula for location reward is calculated by intersection over union of two spans:
    \begin{equation}
    \label{eq_loc_rew}
    r_{location} = \frac{interArea}{(end-start)+(end\_gt-start\_gt)-interArea}
    \end{equation}
where $interArea$ is the length of the intersection between two spans.
    
\subsubsection{Label reward}
The label reward enforces the correctness of the label (field/tag) of the extracted answer. Particularly, given a question $q_t$, if the IE model extracts the answer $\hat{y} = D[start:end+1]$, we expect that the actual label of the span $(start,end)$ in the document $D$ matches to $q_t$, i.e:
    \begin{equation}
    \label{eq_label_rew}
    r_{label} =
    \begin{cases}
      1 & \text{if $label(\hat{y}) = q_t$}\\
      0 & \text{if $label(\hat{y}) \neq q_t$ and $label(\hat{y}) = other$}\\
      -1 & \text{if $label(\hat{y}) \neq q_t$ and $label(\hat{y}) \neq other$}\\
    \end{cases} 
    \end{equation}
where $label()$ returns the actual tag of the answer span $\hat{y}$, by obtaining the majority of token's ground truth tags in this span, $other$ is a special tag we assign for tokens that do not belong to any targeted field. Intuitively, this reward penalizes more severely if the model mis-recognizes an entity as another targeted field rather than its ground truth field.

\subsubsection{Semantic reward}
The semantic reward incentivizes the semantic matching between the output answer and the ground truth one. We utilize the pretrained multilingual sentence encoder \cite{feng2022language} to compute the sentence embeddings of the output and the ground truth, and then measure their Cosine similarity in the latent embedding space.
    \begin{equation}
    \label{eq_semantic_rew}
    r_{semantic} = cosine(enc(y),enc(\hat{y}))
    \end{equation}
where $enc()$ returns the vector of an input by using the sentence encoder.

The final formula for the unified reward.
\begin{equation}
\label{eq_final_rew}
    r = \alpha_1 \times r_{string} + \alpha_2 \times r_{location} + \alpha_3 \times r_{label} + \alpha_4 \times r_{semantic}
\end{equation}
where $\alpha_i$'s are the corresponding weights for each reward component.

\newcommand\Tau{\mathbb{T}}
\begin{algorithm}[!thb]
\caption{Training protocol}
\label{alg_train}
\begin{algorithmic}[1]
\Require Dataset $\mathbb{D}$, Pretrained SpanIE-Recur $\pi_\theta()$, Maximum number of iterations for RL finetuning $n_{max}$.
\State Perform traditional supervised training of $\pi_\theta()$ on $\mathbb{D}$ until convergence. $i \gets 0$
\While{$i \leq n_{max}$} 
    \State Initialize set of trajectories $\Tau \gets \{\}$. Sample set of documents $\{D_k\}$ from $\mathbb{D}$.
    \For{$D_k$ in $\{D_k\}$}
        \State Initialize trajectory $\tau \gets \{\}$, $t \gets 0$.
        \For{$q_t$ in all questions}
            \State The agent is in state $s_t \gets (D_k, q_t)$.
            \State The agent takes action $a_t \sim \pi_\theta(s_t)$ as described in Section \ref{sec:formulation_rl}.
            \State The environment returns corresponding reward $r_t$ calculated by Equation \ref{eq_final_rew}.
            \State The environment returns the timestep for the next state $t \gets t+1$.
            \State $\tau \gets \tau \cup \{s_t, a_t, r_t\}$. $i \gets i+1$
        \EndFor
        \State $\Tau \gets \Tau \cup \tau$
    \EndFor
    \State Given trajectories $\Tau$, calculate PPO loss function by Equation \ref{eq_ppo}.
    \State Backpropagate the gradients and update the parameters of $\pi_\theta$.
\EndWhile
\State \Return $\pi_\theta$
\end{algorithmic}
\end{algorithm}

\subsection{Transfer learning with ground-truth labels}\label{sec:ground-truth}

In this section, we describe the RL fine-tuning procedure based on the following protocol. Starting with a pretrained weight model (e.g. LayoutLM), we finetune the pretrained model on the training set with normal supervised training until convergence. In practice, this training set is divided into smaller subsets with a given size (2\%, 5\%, 10\% or 100\% of the entire training data). We finetune the model further with our reinforcement learning for up to $n_{max}=100000$ iterations. During this process, we utilize the ground truth labels provided with each dataset to compute the aforementioned reward functions. These ground truth labels comprise of the actual texts and their corresponding starting and ending positions in the context. The goal is to observe whether our reinforcement finetuning helps to improve the performance of the normal supervised training of the current state-of-the-art model.

\subsection{Experts' feedback as reward}\label{sec:hitl}

In addition to using ground-truth answers to compute the rewards, we propose to use expert's feedback to train the model. In this setting, we mimic the real environment where human can give feedback to the model via an interactive interface. The interface for feedback is designed as follows. Given a document and a question, the model outputs softmax distributions over starting and ending positions of the possible answers. We then sample from the distributions to get five pairs of start-end actions, representing top five answer candidates, which are equivalent to five options/buttons on the screen. To account for cases when the model cannot provide good candidates, we provide a sixth option - a button named No good options available. Experts could select one of the six options that is most appropriate for the question as feedback to the model. The selected candidate is used as ground truth $y$ to compute the reward using Equation \ref{eq_final_rew} above. Figure \ref{fig:hitl-results} (a) shows the interface of our system.

\subsection{Training algorithm}

We describe the detailed protocol to train the information extractor (SpanIE-Recur) with our rewards in Algorithm \ref{alg_train}. We sample a document and its set of questions from the training data to form the RL trajectory. For each question, we concatenate the document and question representations to build the current state. Then, we sample one start-end action from the softmax layers of SpanIE-Recur to extract the candidate answer for the question. The reward is computed based on the candidate answer and the ground truth to construct the RL's loss function to optimize SpanIE-Recur.

\section{Settings and Evaluation Metrics}\subsubsection{Datasets}
We use two public and two private datasets in our experiments. The two public datasets are in English: SROIE \cite{huang2019icdar2019} and CORD \cite{park2019cord}. SROIE is a collection of scanned receipt images, where each receipt has four fields to extract: \textit{address}, \textit{company}, \textit{date}, \textit{total}. CORD contains receipts collected from Indonesian shops and restaurants, where there are 30 semantic labels defined under 4 categories, such as \textit{store information}, \textit{payment information}, \textit{menu}, \textit{total}. Compared to SROIE, document images in CORD are captured in the wild, thus the data is more noisy and has lower quality. For private datasets, we collect two in-house datasets: the first one is a collection of Japanese technical document images and the second one is a dataset of Japanese invoices. The statistics of the four datasets are listed in Table \ref{tab:stat-data}.
\begin{table}[!htb]
\setlength{\tabcolsep}{5.5pt}
\caption{Statistic of datasets}\label{tab:stat-data}
\centering
\begin{tabular}{lcccc}
\hline
Dataset & \# of fields & train & development & test \\ \hline
SROIE     & 4  & 626 & 0 & 347   \\
CORD      & 30 & 800 & 100 & 100 \\
Inhouse-1 & 12 & 481 & 0 & 149   \\
Inhouse-2 & 12 & 1032 & 0 & 433  \\ \hline
\end{tabular}
\end{table}

\subsubsection{Evaluation metrics}
The results are reported on the test split of the corresponding dataset by evaluating the standard field-level Precision, Recall, and F1 scores (weighted average) as in previous studies \cite{xu2020layoutlm,xu2021layoutlmv2}.

\subsubsection{Baselines}
As far as we know, there is only one prior work \cite{wang2022towards} which targets the same few-shot learning problem on business documents. However, they only target few simple fields in their experiments and do not cover the broad range of datasets. Thus, we consider the supervised learning (SL) result of SpanIE-Recur \cite{nguyen2021span} (with the LayoutLM backbone) as a competitive baseline, which is also used in \cite{wang2022towards}.

\subsubsection{Implementation detail}
Our model uses \textit{LayoutLM-base} with 12 Transformer blocks, the hidden size is 768, the input sequence length is 512 and the total number of parameters is 113M. The question embedding size is also 768. All experiments are conducted on a single Tesla T4 GPU. For all experiments, we use a learning rate of $5e-5$ and the Adam optimizer in the supervised training stage, with the standard cross-entropy loss. The hyper-parameters for PPO training are: the learning rate of $1e-6$, discount factor of 0.95 and clipping rate of 0.2. We set the weights of reward components $\alpha_i$'s to 0.25. These hyper-parameters are manually tuned for the Inhouse-1 dataset and used across experiments. To reduce the effort of tuning in future works, we may use methods that automatically learn hyper-parameter scheduling \cite{le2022episodic}. 

\section{Experimental Results}We confirm the efficiency of our approach in two settings: transfer learning with limited data by using ground-truth labels and learning with humans' feedback. While the former uses the ground-truth answer to construct different rewards for training the model (Section \ref{sec:ground-truth}), the latter uses human's feedback (see Section \ref{sec:hitl}). We also conduct ablation studies to verify the contribution of each reward type.

\begin{table}[!t]
\setlength{\tabcolsep}{4.5pt}
\caption{Results of transfer learning on public datasets.}
\label{tab:transfer_pub}
\centering
\begin{tabular}{ll|ccc|cccc}
\hline
\textbf{}                                            & \textbf{}      & \multicolumn{3}{c|}{\textbf{Original SL}} & \multicolumn{4}{c}{\textbf{+RL}}                  \\ \hline
\multicolumn{2}{c|}{\textbf{Dataset}} &
  \multicolumn{1}{c}{\textbf{Precision}} &
  \multicolumn{1}{c}{\textbf{Recall}} &
  \multicolumn{1}{c|}{\textbf{F1}} &
  \multicolumn{1}{c}{\textbf{Precision}} &
  \multicolumn{1}{c}{\textbf{Recall}} &
  \multicolumn{1}{c|}{\textbf{F1}} &
  \multicolumn{1}{c}{$\Delta$\textbf{F1}} \\ \hline
\multicolumn{1}{l|}{\multirow{4}{*}{\textbf{CORD}}}  & \textbf{2\%}   & 21.69     & 2.77      & 3.04     & 40.11 & 15.94 & \multicolumn{1}{r|}{21.23} & 18.19 \\
\multicolumn{1}{l|}{}                                & \textbf{5\%}   & 64.33     & 47.83     & 51.48    & 66.39 & 52.69 & \multicolumn{1}{r|}{57.83} & 6.35  \\
\multicolumn{1}{l|}{}                                & \textbf{10\%}  & 79.45     & 72.98     & 75.32    & 80.76 & 73.05 & \multicolumn{1}{r|}{76.12} & 0.80  \\
\multicolumn{1}{l|}{}                                & \textbf{100\%} & 96.31     & 94.91     & 95.39    & 97.11 & 95.03 & \multicolumn{1}{r|}{95.71} & 0.32  \\ \hline
\multicolumn{1}{l|}{\multirow{4}{*}{\textbf{SROIE}}} & \textbf{2\%}   & 80.93     & 78.08     & 79.46    & 80.61 & 79.77 & \multicolumn{1}{r|}{80.16} & 0.70   \\
\multicolumn{1}{l|}{}                                & \textbf{5\%}   & 85.94     & 80.59     & 83.09    & 86.07 & 82.09 & \multicolumn{1}{r|}{83.95} & 0.86  \\
\multicolumn{1}{l|}{}                                & \textbf{10\%}  & 88.33     & 85.27     & 86.74    & 89.22 & 85.37 & \multicolumn{1}{r|}{87.20} & 0.46  \\
\multicolumn{1}{l|}{}                                & \textbf{100\%} & 91.83     & 91.45     & 91.61    & 91.89 & 91.54 & \multicolumn{1}{r|}{91.68} & 0.07  \\ \hline
\end{tabular}

\caption{Results of transfer learning on our in-house datasets. \textbf{Pretrained} means the in-house pretrained weight we use to initialize the model before performing transfer learning.}
\label{tab:transfer_private}
\centering
\begin{tabular}{ll|ccc|cccc}
\hline
\textbf{}                                                & \textbf{}           & \multicolumn{3}{c|}{\textbf{Original SL}} & \multicolumn{4}{c}{\textbf{+RL}}                  \\ \hline
\multicolumn{2}{c|}{\textbf{Dataset}} &
  \multicolumn{1}{c}{\textbf{Precision}} &
  \multicolumn{1}{c}{\textbf{Recall}} &
  \multicolumn{1}{c|}{\textbf{F1}} &
  \multicolumn{1}{c}{\textbf{Precision}} &
  \multicolumn{1}{c}{\textbf{Recall}} &
  \multicolumn{1}{c|}{\textbf{F1}} &
  \multicolumn{1}{c}{$\Delta$\textbf{F1}} \\ \hline
\multicolumn{1}{l|}{\multirow{5}{*}{\textbf{Inhouse-1}}} & \textbf{Pretrained} & 38.24     & 14.62     & 18.02    & -     & -     & \multicolumn{1}{c|}{-}     & -     \\
\multicolumn{1}{l|}{}                                    & \textbf{2\%}        & 74.44     & 56.63     & 62.08    & 69.39 & 65.17 & \multicolumn{1}{r|}{66.35} & 4.27  \\
\multicolumn{1}{l|}{}                                    & \textbf{5\%}        & 86.45     & 76.64     & 80.35    & 84.15 & 75.60 & \multicolumn{1}{r|}{79.63} & -0.72 \\
\multicolumn{1}{l|}{}                                    & \textbf{10\%}       & 89.91     & 80.29     & 83.13    & 89.11 & 82.92 & \multicolumn{1}{r|}{85.75} & 2.62  \\
\multicolumn{1}{l|}{}                                    & \textbf{100\%}      & 94.34     & 94.02     & 94.03    & 94.36 & 94.98 & \multicolumn{1}{r|}{94.37} & 0.34  \\ \hline
\multicolumn{1}{l|}{\multirow{5}{*}{\textbf{Inhouse-2}}} & \textbf{Pretrained} & 73.83     & 40.53     & 48.06    & -     & -     & \multicolumn{1}{c|}{-}     & -     \\
\multicolumn{1}{l|}{}                                    & \textbf{2\%}        & 82.18     & 68.74     & 73.71    & 81.18 & 69.62 & \multicolumn{1}{r|}{74.96} & 1.25  \\
\multicolumn{1}{l|}{}                                    & \textbf{5\%}        & 85.94     & 76.32     & 79.42    & 83.09 & 78.95 & \multicolumn{1}{r|}{80.49} & 1.07  \\
\multicolumn{1}{l|}{}                                    & \textbf{10\%}       & 88.27     & 81.21     & 82.81    & 88.25 & 82.17 & \multicolumn{1}{r|}{82.91} & 0.10  \\
\multicolumn{1}{l|}{}                                    & \textbf{100\%}      & 91.42     & 91.30     & 91.35    & 91.42 & 91.30 & \multicolumn{1}{r|}{91.35} & 0.00  \\ \hline
\end{tabular}
\end{table}


\subsection{Transfer learning with limited data}\label{pubdata}

\subsubsection{Training procedure}\label{sec_transfer}
We conduct experiments from different subsets of the training data to show the benefit of our proposed reinforcement finetuning mechanism. For the public datasets, we use the pretrained LayoutLM weight \emph{layoutxlm-no-visual}.\footnote{\url{https://huggingface.co/taprosoft/layoutxlm-no-visual}} We use an in-house pretrained weight to initialize the model for the private datasets. 
  
\subsubsection{Benchmarking results}

We report the results on CORD and SOIRE datasets in Table \ref{tab:transfer_pub}. Overall, we can observe clearer improvements over the supervised baseline.
RL finetuning makes the marginal improvements in the 10\% and 100\% data scenarios. The contribution is more remarkable in the limited data regimes, where on the CORD dataset, we observe the increase of 18.19\% and 6.35\% of F1 score for 2\% and 5\% training data, respectively. On SROIE, the corresponding improvements are 0.70\% and 0.86\%. We hypothesize that with large enough amount of data, the backbone model trained with SL receives abundant training signal to become a strong backbone, but in limted data scenarios, RL performs better since it can exploit different aspects of training signals. Since CORD dataset has much more fields than SROIE, it is harder for SL models to learn in limited data scenarios on CORD, thus leaving rooms for RL contribution. This is the rationale behind the large improvements on CORD compared to SROIE dataset.

Table \ref{tab:transfer_private} shows the results on our in-house datasets. We observe similar behaviors of the RL-finetuned models. In most cases, RL finetuning can improve the performances, noticeably when the performance of SL counterparts are humble in 2\% data scenarios (4.27\% and 1.25\% F1 score improvements on Inhouse-1 and Inhouse-2 respectively).

\subsubsection{Ablation Study}

In this section, we investigate the impact of different reward components in the unified reward function. We use the setting with the 2\% training data scenario, but we ablate each single reward component respectively. Table \ref{tab:ablation} reports the F1 score of each ablated model on CORD and SOIRE data. Without string matching and location rewards, the performances drop significantly, but retain the improvements upon the SL counterpart. We observe the most severe drop when removing the semantic reward component on both datasets, which emphasizes the importance of the semantic matching between output and ground truth answers. Ablating the label reward on CORD dataset results in comparable performance to that of the full unified reward, but on SROIE dataset, we observe a large drop, thus confirms the helpfulness of the label reward component.

\begin{table}[!t]
\setlength{\tabcolsep}{3.5pt}
\caption{Effect of ablating different reward components.}
\label{tab:ablation}
\centering
\begin{tabular}{l|cccc|cccc}
\hline
 & 
  \multicolumn{4}{c|}{\textbf{CORD}} & \multicolumn{4}{c}{\textbf{SROIE}}
  \\ \hline
  \textbf{Model} &
  \multicolumn{1}{c}{\textbf{Precision}} &
  \multicolumn{1}{c}{\textbf{Recall}} &
  \multicolumn{1}{c}{\textbf{F1}} &
  \multicolumn{1}{c|}{$\Delta$\textbf{F1}} &
  \multicolumn{1}{c}{\textbf{Precision}} &
  \multicolumn{1}{c}{\textbf{Recall}} &
  \multicolumn{1}{c}{\textbf{F1}} &
  \multicolumn{1}{c}{$\Delta$\textbf{F1}}
  \\ \hline
  \textbf{SL model}            & 21.69 & 2.77  & 3.04  & \multicolumn{1}{c|}{-} & 80.93 & 78.08 & 79.46 & \multicolumn{1}{c}{-}     \\
  \textbf{w/o string matching} & 37.87 & 5.46  & 7.79  & 4.75 & 80.91 & 78.17 & 79.67 & 0.21 \\
  \textbf{w/o location}        & 47.07 & 5.16  & 7.82  & 4.78 & 80.42 & 78.85 & 79.83 & 0.37 \\
  \textbf{w/o label}           & 31.19 & 17.51 & 2.22  & 19.16 & 80.52 & 78.75 & 79.61 & 0.15 \\
  \textbf{w/o semantic}        & 2.27  & 2.54  & 2.40  & -0.64 & 81.47 & 77.69 & 79.52 & 0.06 \\
  \textbf{Full unified reward} & 40.11 & 15.94 & 21.23 & 18.19 & 80.61 & 79.77 & 80.16 & 0.70 \\ \hline
\end{tabular}
\end{table}

\subsection{Learning with experts' feedback}
\begin{figure}
	\begin{minipage}[c]{0.5\linewidth}
		\centering
		\includegraphics[width=\linewidth]{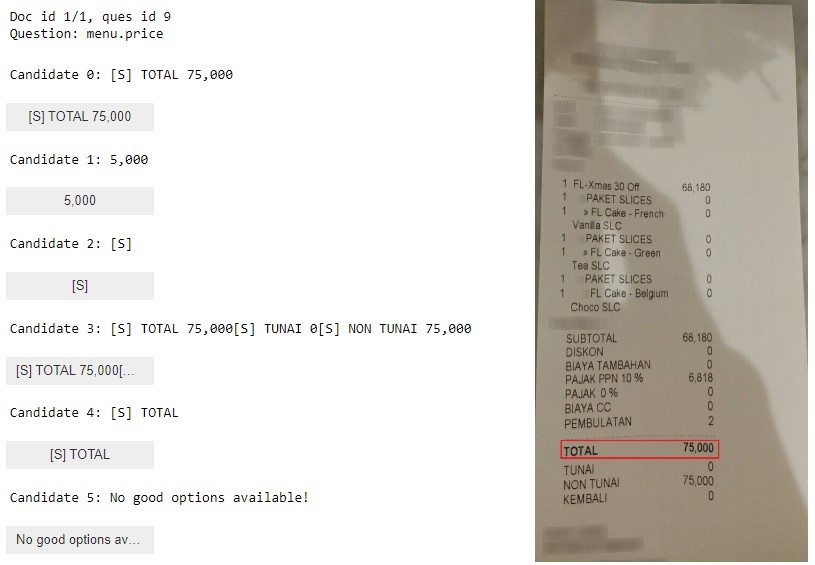}\\
		(a) Example of interaction interface
	\end{minipage}\hfill
	\begin{minipage}[c]{0.5\linewidth}
		\centering
		\includegraphics[width=\linewidth]{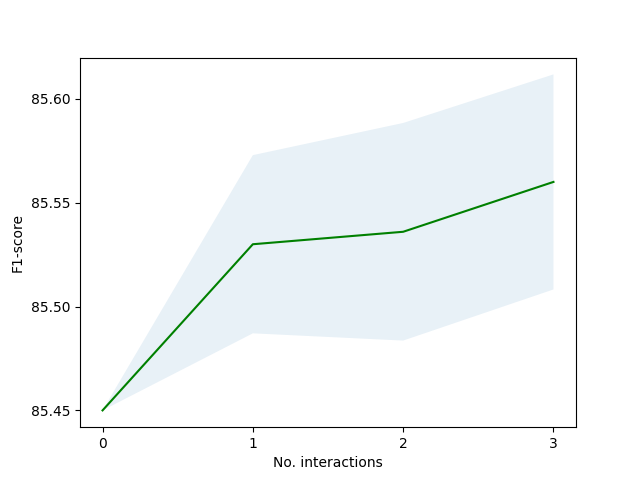}\\
		(b) CORD: F1 score with human feedback
	\end{minipage}
	\caption{Learning with experts' feedback.}
	\label{fig:hitl-results}
\end{figure}

We conduct an interactive experiment that allows a domain expert to train the system with preference reward via feedback. For this experiment, we sample five sets with four documents each from the dev set of CORD, use them to get feedback from experts, and evaluate the results on the test set after each interaction. Figure \ref{fig:hitl-results}(b) shows the F1-score with the mean and standard deviation for five sets, with the result at interaction 0 being from the backbone using only supervised learning on 100\% of the training set. After three interactions, the average F1-score of five sets increases marginally at around 0.11\%. Yet, considering the small number of documents sampled for feedback and a decent performance of the backbone at the beginning, this increase is notable.

\section{Discussion}
This work proposes a novel end-to-end reinforcement learning model with task-focused reward for document image extraction task. Our experiments show that for both English and Japanese documents, our model outputs competitive results with the traditional supervised learning approach when having full access to training data, while performing significantly better when reducing the amount of training data, notably the 18.19\% improvement in F1-score for 2\% of the CORD training set. In addition, we provide an interactive interface session to get feedback from experts to improve the model’s performances. Our results suggest that with as little as two to three interactions, users are able to witness the changes in output for the betterment. Despite this, the feedback interface, along with its connection to the RL model, is still in preliminary state. For future works, one should consider not only the selected candidate but also the non-selected ones, and how to represent them in the reward model.


\printbibliography

\end{document}